\documentclass[twoside,twocolumn,9pt]{article}
\usepackage{extsizes}
\usepackage[super,sort&compress,comma]{natbib} 
\usepackage[version=3]{mhchem}
\usepackage[left=1.5cm, right=1.5cm, top=1.785cm, bottom=2.0cm]{geometry}
\usepackage{balance}
\usepackage{mathptmx}
\usepackage{sectsty}
\usepackage{graphicx} 
\usepackage{lastpage}
\usepackage[format=plain,justification=justified,singlelinecheck=false,font={stretch=1.125,small,sf},labelfont=bf,labelsep=space]{caption}
\usepackage{float}
\usepackage{fancyhdr}
\usepackage{fnpos}
\usepackage[english]{babel}
\addto{\captionsenglish}{%
  
}
\usepackage{array}
\usepackage{droidsans}
\usepackage{charter}
\usepackage[T1]{fontenc}
\usepackage[usenames,dvipsnames]{xcolor}
\usepackage{setspace}
\usepackage[compact]{titlesec}
\usepackage{hyperref}

\usepackage{epstopdf}

\usepackage{adjustbox}
\usepackage{multirow}
\usepackage{booktabs}
\usepackage[normalem]{ulem}
\useunder{\uline}{\ul}{}

\definecolor{cream}{RGB}{222,217,201}

\begin{document}

\pagestyle{fancy}
\thispagestyle{plain}
\fancypagestyle{plain}{
\renewcommand{\headrulewidth}{0pt}
}

\makeFNbottom
\makeatletter
\renewcommand\LARGE{\@setfontsize\LARGE{15pt}{17}}
\renewcommand\Large{\@setfontsize\Large{12pt}{14}}
\renewcommand\large{\@setfontsize\large{10pt}{12}}
\renewcommand\footnotesize{\@setfontsize\footnotesize{7pt}{10}}
\makeatother

\renewcommand{\thefootnote}{\fnsymbol{footnote}}
\renewcommand\footnoterule{\vspace*{1pt}%
\color{cream}\hrule width 3.5in height 0.4pt \color{black}\vspace*{5pt}} 
\setcounter{secnumdepth}{5}

\makeatletter 
\renewcommand\@biblabel[1]{#1}            
\renewcommand\@makefntext[1]%
{\noindent\makebox[0pt][r]{\@thefnmark\,}#1}
\makeatother 
\renewcommand{\figurename}{\small{Fig.}~}
\sectionfont{\sffamily\Large}
\subsectionfont{\normalsize}
\subsubsectionfont{\bf}
\setstretch{1.125} 
\setlength{\skip\footins}{0.8cm}
\setlength{\footnotesep}{0.25cm}
\setlength{\jot}{10pt}
\titlespacing*{\section}{0pt}{4pt}{4pt}
\titlespacing*{\subsection}{0pt}{15pt}{1pt}

\fancyfoot{}
\fancyfoot[LO,RE]{\vspace{-7.1pt}\includegraphics[height=9pt]{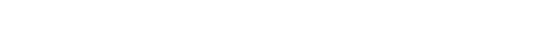}}
\fancyfoot[CO]{\vspace{-7.1pt}\hspace{11.9cm}\includegraphics{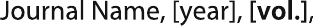}}
\fancyfoot[CE]{\vspace{-7.2pt}\hspace{-13.2cm}\includegraphics{head_foot/RF}}
\fancyfoot[RO]{\footnotesize{\sffamily{1--\pageref{LastPage} ~\textbar  \hspace{2pt}\thepage}}}
\fancyfoot[LE]{\footnotesize{\sffamily{\thepage~\textbar\hspace{4.65cm} 1--\pageref{LastPage}}}}
\fancyhead{}
\renewcommand{\headrulewidth}{0pt} 
\renewcommand{\footrulewidth}{0pt}
\setlength{\arrayrulewidth}{1pt}
\setlength{\columnsep}{6.5mm}
\setlength\bibsep{1pt}

\makeatletter 
\newlength{\figrulesep} 
\setlength{\figrulesep}{0.5\textfloatsep} 

\newcommand{\topfigrule}{\vspace*{-1pt}%
\noindent{\color{cream}\rule[-\figrulesep]{\columnwidth}{1.5pt}} }

\newcommand{\botfigrule}{\vspace*{-2pt}%
\noindent{\color{cream}\rule[\figrulesep]{\columnwidth}{1.5pt}} }

\newcommand{\dblfigrule}{\vspace*{-1pt}%
\noindent{\color{cream}\rule[-\figrulesep]{\textwidth}{1.5pt}} }

\makeatother

\twocolumn[
  \begin{@twocolumnfalse}
{\includegraphics[height=30pt]{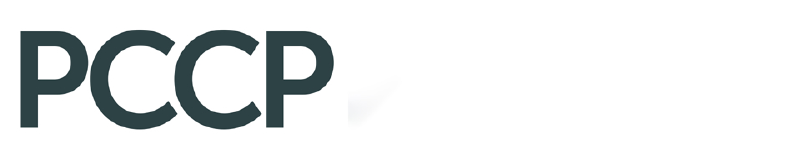}\hfill\raisebox{0pt}[0pt][0pt]{\includegraphics[height=55pt]{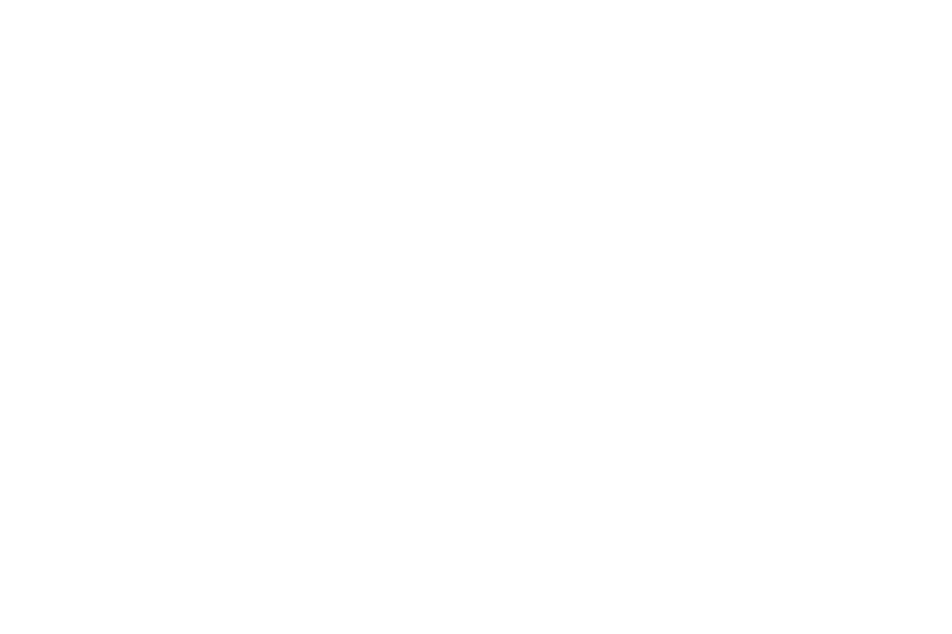}}\\[1ex]
\includegraphics[width=18.5cm]{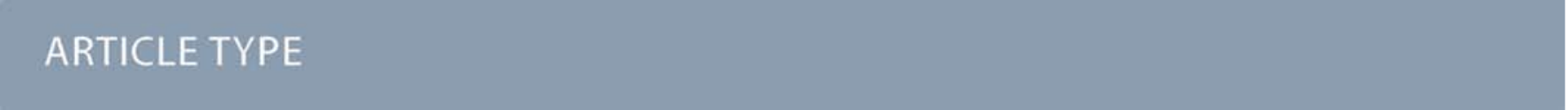}}\par
\vspace{1em}
\sffamily
\begin{tabular}{m{4.5cm} p{13.5cm} }

\includegraphics{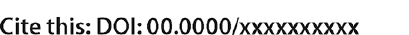} & \noindent\LARGE{\textbf{The Kinetic Energy of PAH Dication and Trication \mbox{Dissociation} Determined by Recoil-Frame Covariance Map Imaging}} \\
\vspace{0.3cm} & \vspace{0.3cm} \\

 & \noindent\large{Jason W. L. Lee,\textit{$^{a,b}$} 
 Denis S. Tikhonov,\textit{$^{a,c}$}
 Felix Allum,\textit{$^{b}$} 
 Rebecca Boll,\textit{$^{d}$}
 Pragya Chopra,\textit{$^{a,c}$} 
 Benjamin Erk,\textit{$^{a}$} 
 Sebastian Gruet,\textit{$^{a}$} 
 Lanhai He,\textit{$^{a}$} 
 David Heathcote,\textit{$^{b}$} 
 Mehdi M. Kazemi,\textit{$^{a}$} 
 Jan Lahl,\textit{$^{e}$} 
 Alexander K. Lemmens,\textit{$^{f,g}$} 
 Donatella Loru,\textit{$^{a}$} 
 Sylvain Maclot,\textit{$^{h,i}$}
 Robert Mason,\textit{$^{b}$} 
 Erland Müller,\textit{$^{a}$} 
 Terry Mullins,\textit{$^{j}$} 
 Christopher Passow,\textit{$^{a}$} 
 Jasper Peschel,\textit{$^{e}$} 
 Daniel Ramm,\textit{$^{a}$} 
 Amanda L. Steber,\textit{$^{a,c,k}$}
 Sadia Bari,\textit{$^{a}$} 
 Mark Brouard,\textit{$^{b}$} 
 Michael Burt,\textit{$^{b}$} 
 Jochen Küpper,\textit{$^{l,k,j}$} 
 Per Eng-Johnsson,\textit{$^{e}$} 
 Anouk M. Rijs,\textit{$^{f}$} 
 Daniel Rolles,\textit{$^{m}$} 
 Claire Vallance,\textit{$^{b}$} 
 Bastian Manschwetus,\textit{$^{a}$} 
 and Melanie Schnell\textit{$^{a}$}} \\

\includegraphics{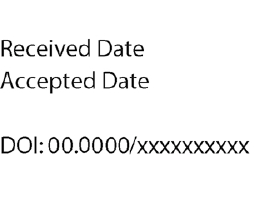} & \noindent\normalsize{We investigated the dissociation of dications and trications of three polycyclic aromatic hydrocarbons (PAHs), fluorene, phenanthrene, and pyrene. PAHs are a family of molecules ubiquitous in space and involved in much of the chemistry of the interstellar medium. In our experiments, ions are formed by interaction with 30.3 nm extreme ultraviolet (XUV) photons, and their velocity map images are recorded using a PImMS2 multi-mass imaging sensor. Application of recoil-frame covariance analysis allows the total kinetic energy release (TKER) associated with multiple fragmentation channels to be determined to high precision, ranging 1.94-2.60 eV and 2.95-5.29 eV for the dications and trications, respectively. Experimental measurements are supported by Born-Oppenheimer molecular dynamics (BOMD) simulations.} \\

\end{tabular}

 \end{@twocolumnfalse} \vspace{0.6cm}

  ]

\renewcommand*\rmdefault{bch}\normalfont\upshape
\rmfamily
\section*{}
\vspace{-1cm}


\footnotetext{\textit{$^{a}$~Deutsches Elektronen-Synchrotron DESY, Germany}}
\footnotetext{\textit{$^{b}$~Department of Chemistry, University of Oxford, United Kingdom}}
\footnotetext{\textit{$^{c}$~Institute of Physical Chemistry, Christian-Albrechts-Universität zu Kiel, Germany}}
\footnotetext{\textit{$^{d}$~European XFEL, Germany}}
\footnotetext{\textit{$^{e}$~Department of Physics, Lund University, Sweden}}
\footnotetext{\textit{$^{f}$~Radboud University, FELIX Laboratory, The Netherlands}}
\footnotetext{\textit{$^{g}$~Van’t Hoff Institute for Molecular Sciences, University of Amsterdam}}
\footnotetext{\textit{$^{h}$~KTH Royal Institute of Technology, Sweden}}
\footnotetext{\textit{$^{i}$~Physics Department, University of Gothenburg, Sweden}}
\footnotetext{\textit{$^{j}$~Department of Physics, Universität Hamburg, Germany}}
\footnotetext{\textit{$^{k}$~Center for Ultrafast Imaging, Universität Hamburg, Germany}}
\footnotetext{\textit{$^{l}$~Center for Free-Electron Laser Science CFEL, Deutsches Elektronen-Synchrotron DESY, Germany}}
\footnotetext{\textit{$^{m}$~J.R. Macdonald Laboratory, Department of Physics, Kansas State University, KS, USA}}


\footnotetext{\dag~Electronic Supplementary Information (ESI) available. See DOI: 10.1039/cXCP00000x/}




\section{Introduction}
Polycyclic aromatic hydrocarbons (PAHs) are significant components of the interstellar medium (ISM) with infrared emission measurements estimating that 10-15\% of total galactic carbon is contained in PAHs. \cite{joblin201125, lagache2004polycyclic} Given their prevalence, PAH molecules are often cited as potential carriers of the Diffuse Interstellar Bands (DIBs), a set of over 500 absorption bands that were first identified in 1922, \cite{heger1922further} of which only one carrier (C$_{60}^+)$ has currently been identified. \cite{campbell2015laboratory} Within the ISM, PAHs are exposed to high energy photons from extraterrestrial sources, inducing ionization, isomerization, and/or fragmentation. The abundance of PAHs means that such processes will significantly influence interstellar chemistry; consequently, investigation into PAHs has motivated laboratory experiments for many decades. Improving our understanding of the species created in photodissociative events and their formation mechanisms provides a valuable insight into the evolution of the ISM.

\begin{figure*}
\centering
\includegraphics[scale=0.65]{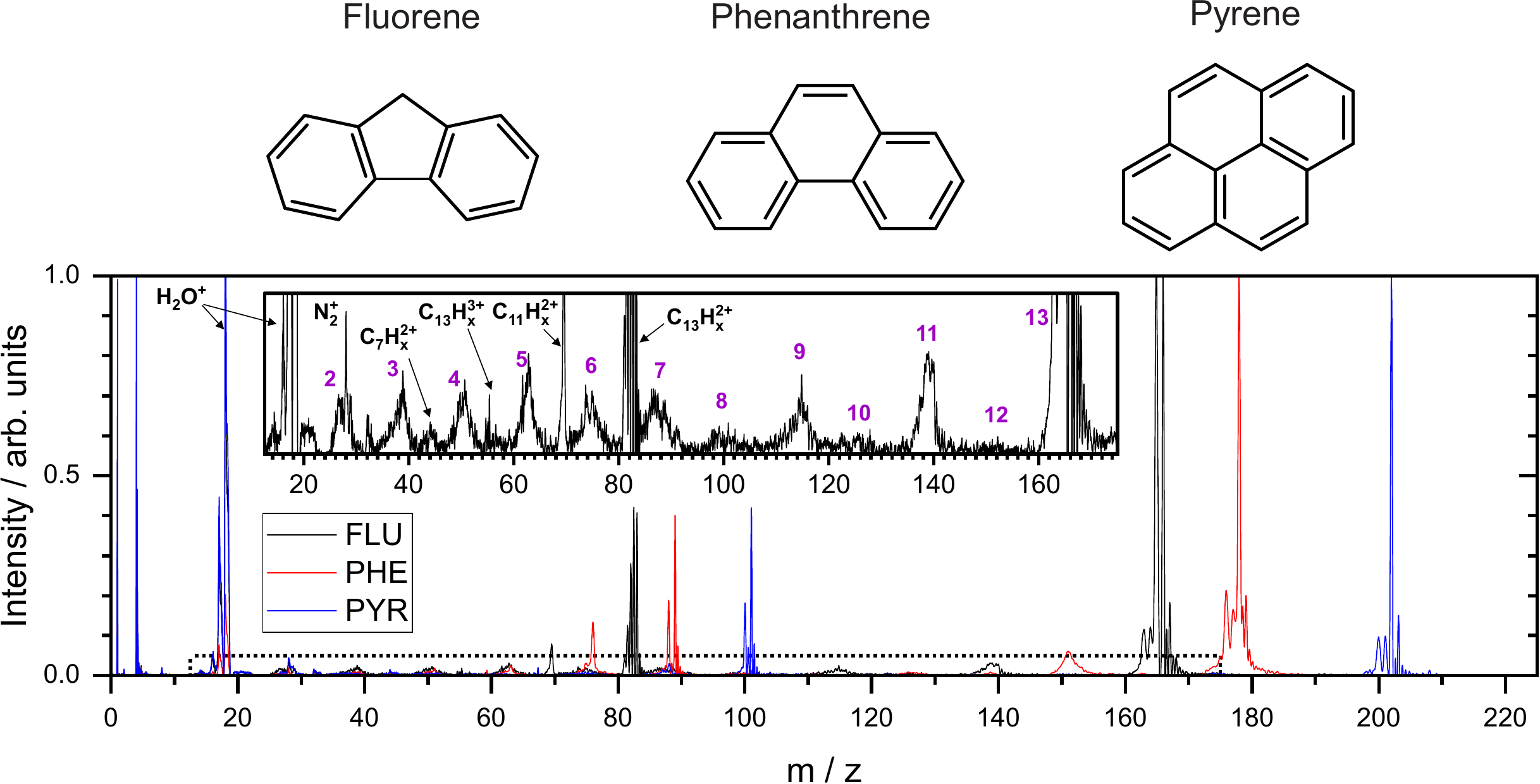}
\caption{\label{fig:mass-spectrum} Top: molecular structures of the three molecules studied in this report. Bottom: the normalized mass spectra recorded following interaction with the 40.9 eV photons. Inset is a 20x expanded view of the mass spectrum for FLU, region denoted by the dotted line on the main spectrum. Purple numbers indicate the number of carbon atoms in the monocations primarily associated with the broad peaks in the mass spectrum.}
\end{figure*}

Reported in 1970, Beynon \textit{et al}.~measured the dissociation energies, or total kinetic energy releases (TKERs), of a small series of aromatic molecules, ranging from benzene (C$_6$H$_6$) to anthracene (C$_{14}$H$_{10}$). \cite{beynon1970ion} Their experiment recorded ion kinetic energy spectra (IKES) following dissociation ionization of the molecule by electron impact at 80 eV. Dication dissociation TKER values in the range 1.2-2.9 eV were reported, depending on the molecule and the fragmentation products. A follow up paper with the same experimental setup found the TKER of triply-charged biphenyl to be 4.5 eV. \cite{beynon1970decomposition} The authors found that a simple model of two charges on the original structure of biphenyl predicted a much higher dissociation energy value due to the Coulombic interaction, and therefore concluded that ring-opened structures, where the charges could be separated by a greater distance, were more consistent.

Further electron impact studies on PAHs at 70 eV were later performed by Kingston \textit{et al}.~using the improved measurement technique of mass-analyzed ion kinetic energy spectra (MIKES). \cite{beynon1973design, kingston1985multiple} The dataset measured MIKES for aromatic molecules with one to four six-membered rings. The dication and trication TKER values were reported in the range 1.9-2.7 eV and 3.9-5.5 eV, respectively, and were rationalized to be consistent with ring-opened structures, similar to Beynon \textit{et al}.

Triply-charged PAHs with four to six six-membered rings were studied by March \textit{et al}.~and Koyanagi \textit{et al}.~using MIKES combined with 70 eV and 100 eV electron impact, respectively. The former assessed the trication TKER of benzo[a]pyrene (C$_{20}$H$_{12}$) to be \textasciitilde 4.75 eV, and the latter recorded energies of \mbox{4.3-5.7 eV} for the PAHs investigated. Interestingly, their modelling found that these values could occur from the intact molecular skeleton, indicating that the larger PAHs in these studies did not undergo ring-opening prior to dissociation.


The interaction of PAHs with vacuum ultraviolet (VUV) and extreme ultraviolet (XUV) photons was originally reported by Eland \textit{et al}.~through the use of rare gas lamps and photoelectron-photoion-photoion coincidence (PEPIPICO) techniques. \cite{eland1989new, leach1989formation, leach1989formation2} One experiment investigated fully deuterated naphthalene (NAP-d8, C$_{10}$D$_8$) with photon energies of 34.8 eV and 40.8 eV. TKERs in the range 0.5-2.5 eV were reported for the formation of two monocations from the parent dication. In agreement with earlier electron impact studies on small PAHs, the authors hypothesized significant molecular rearrangement before dissociation to reduce the Coulombic interaction sufficiently to yield the TKER values recorded. Studies by Reitsma \textit{et al}.~generated the parent dication of naphthalene (NAP) by collision with 30 keV He$^{2+}$ ions. \cite{reitsma2013ion} Analysis of the fragment ions from the NAP dication dissociation were analyzed to provide TKER values in the range 2-3 eV, slightly higher than the Eland values. Density functional theory (DFT) calculations were performed to explore possible transition states and dissociation channels, calculating activation barriers and reverse barriers, showing some consistency with the shape of the TKER distributions obtained.



There have additionally been a wide number of theoretical studies on PAH dissociation, aimed at understanding the cation stability and dissociation dynamics. Malloci \textit{et al}.~presented a DFT study of 40 PAH dications ranging from azulene (C$_{10}$H$_8$) to circumvalene (C$_{60}$H$_{20}$), calculating the adiabatic double ionization energies and photo-absorption cross-section up to 30 eV. \cite{malloci2007theoretical} They concluded that the radiative environment of the H I regions (i.e.~areas of the ISM composed primarily of neutral atomic hydrogen) should efficiently generate the dication for all species investigated. Holm \textit{et al}.~also performed in-depth DFT calculations on five small PAH molecules, determining that at a charge state above the dication, small catacondensed PAHs (e.g.~naphthalene, biphenylene and anthracene) become thermodynamically unstable with respect to dissociation. \cite{holm2011dissociation} Theoretical work performed by Simon \textit{et al}.~employed self-consistent charge density functional tight-binding ((SCC)-DFTB) method to simulate the dissociation of some PAH monocations. \cite{simon2017dissociation} They predict that the pyrene monocation with 22 eV of internal energy is largely stable, whereas with 30 eV of internal energy, all pyrene molecules are predicted to undergo rapid fragmentation. The stability of pericondensed PAHs, such as pyrene, is matched in the recent publication by West \textit{et al}., which theorizes that the molecule needs enough internal energy to form particular intermediate structures prior to dissociation. \cite{west2019large}

In the present study, three small PAH molecules, fluorene (FLU, C$_{13}$H$_{10}$), pyrene (PYR, C$_{16}$H$_{10}$), and phenanthrene (PHE, C$_{14}$H$_{10}$), are investigated. The structures of these molecules are shown in Fig. \ref{fig:mass-spectrum}. The dication (PAH$^{2+}$) and trication (PAH$^{3+}$) of each molecule are generated using free-electron laser (FEL) radiation of 30.3 nm (40.9 eV), matching the photon energy of the He II line found in the ISM. In addition to the ion mass spectrum, multi-mass velocity-map imaging (VMI) is used to record the velocity distributions of all the resulting fragment ions simultaneously. From this dataset, dissociative pathways from PAH$^{2+}$ and PAH$^{3+}$ are isolated using recoil-frame covariance analysis. \cite{bull2014account, burt2018communication, amini2017alignment, allum2018coulomb, slater2015coulomb, slater2014covariance, lee2020three} Fitting the resulting covariance map images allows an accurate determination of the magnitude of the momentum of each fragment ion and, by extension, the TKER of the dication and trication dissociation processes. The results are supported by Born-Oppenheimer molecular dynamics (BOMD) theoretical simulations.

\section{Methods}

\subsection{Experimental}
These experiments were performed using the CAMP endstation at beamline BL1 at the FLASH free-electron laser. \cite{erk2018camp, feldhaus2010flash} The sample molecules FLU (C$_{13}$H$_{10}$, melting point (mp) = 116 \textdegree{}C), PHE (C$_{14}$H$_{10}$, mp = 101 \textdegree{}C), and PYR (C$_{16}$H$_{10}$, mp = 145 \textdegree{}C) were purchased from Sigma-Aldrich with 98\% purity and used without further purification. The samples were placed in an in-vacuum reservoir and heated to approximately 220–230 \textdegree{}C. Using helium as a carrier gas (1.5-2 bar backing pressure), the molecules were then introduced into ultra-high vacuum (UHV) \textit{via} a supersonic expansion produced by an Even-Lavie high-temperature pulsed valve with opening times of a few tens of microseconds. \cite{even2015even} The resulting molecular beam was skimmed twice to yield well-collimated pulses of isolated PAH molecules. 

The data set in this work was originally recorded for a recently published XUV-IR pump-probe study. \cite{lee2021time} The data presented in this study focuses on the dissociation of the PAH dications and trications which are almost exclusively formed by the XUV pulse. Two important characteristics were found during analysis: firstly, the energy imparted to the moleculse by the IR pulse is relatively weak compared to the XUV pulse, so the IR pulse is able to initiate single ionization and fragmentation, but formation of the dication/trication is negligibly small; and secondly, very similar recoil-frame covariance images and dissociation energies are obtained by analysis of the data sets employing only all XUV pulse compared to data sets employing both the XUV and IR pulses. Consequently, the data set analyzed in this paper involves both the XUV and IR laser pulses aggregated over the pulse delays in order to improve the statistics in calculating the recoil-frame covariance images.

The FEL was tuned to emit 30.3 nm (40.9 $\pm$ 0.4 eV) XUV pulses, estimated to be 90 fs FWHM with a pulse energy of 3.3 $\mu$J (after filter attenuation). A Ti:sapphire laser produced NIR pulses at a central wavelength of 810 nm and a pulse duration of \mbox{60 fs} FWHM. Measurement of the IR laser pulse and focal spot size prior to the beam entering the chamber provided an estimated intensity of 1 $\times$ $10^{13}$ W/cm$^2$ at the interaction region. Temporal overlap between the laser beams (t$_0$) was determined by the appearance of electron sidebands from the helium carrier gas, and the delay between the lasers was evenly distributed in the approximate range $\pm$1 ps. \cite{lee2021time}

The resulting ions were focused by a velocity-map imaging (VMI) spectrometer onto a pair of chevron-stacked microchannel plates (MCPs) coupled to a P47 phosphor screen. \cite{parker1997photoelectron, eppink1997velocity} The ion images from the phosphor screen were captured by a Pixel Imaging Mass Spectrometry 2 (PImMS2) multi-mass imaging sensor housed within a PImMS camera. \cite{clark2012multimass} The PImMS2 sensor records an \textit{x}, \textit{y}, and \textit{t} coordinate for each ion, allowing a time-of-flight (TOF) spectrum and a two dimensional projection of the three dimensional velocity distribution to be acquired for all fragment ions on each laser shot. In addition, higher resolution TOF spectra were recorded by coupling the voltage drop at the back side of the MCP with a 2 GHz ADC (ADQ412AC-4GMTCA from Teledyne SP Devices) through a resistor-capacitor circuit.

\subsection{Recoil Frame Covariance Analysis}

Following separation by arrival time, correlations between the recorded ion images are determined using recoil-frame covariance analysis. This analysis method has been described in a number of previous articles. \cite{minion2022predicting, burt2018communication, amini2017alignment, allum2018coulomb, slater2015coulomb, slater2014covariance, lee2020three} Briefly, the covariance between two variables, $A$ and $B$, is defined as the product of their deviations from their mean values:

\begin{equation*}
\text{cov}(A,B) = \langle(A - \langle A \rangle) \times (B - \langle B \rangle) \rangle = \langle AB \rangle - \langle A \rangle \langle B \rangle
\end{equation*}

where $\langle \ldots \rangle$ indicates a mean value. For the purposes of applying the technique to ion images, $A$ and $B$ are the ion velocities and the signal for ion $A$ is rotated relative to the position of ion $B$. Consequently, the "reference direction" is defined as the velocity vector of the $B$ ions and the signal of $A$ in the covariance map images is shown relative to this reference direction. In the simplest case of a two-body dissociation ($M \rightarrow A + B$), due to conservation of momentum, $A$ and $B$ will always be directly opposed. Therefore, if such a dissociation pathway exists, signal will appear directly opposite the reference direction.

The pixel coordinates of the ion images and covariance map images were converted to momentum by modelling the instrument in SIMION 8.1. \cite{dahl1990simion}

\subsection{Momentum Profile Fitting}

The ion images sample an ensemble of molecules that are oriented differently relative to the detector plane. In typical VMI studies, an Abel-inversion would be applied to account for this, however, the images produced through the recoil-frame covariance analysis lack cyclindrical symmetry, preventing the transformation from being applied. Therefore, the extracted momentum profiles from the dication and trication dissociation were fit using the BiGaussian function in OriginPro 2020 to account for the asymmetry. This function fits a single peak with two half-Gaussians with different width on each side of the centre of the peak according to the following:
$$
y(x) = \begin{cases}
 H \cdot \exp\left(-\frac{(x-x_c)^2}{2 w_1^2} \right) \ , & x < x_c \\
H \cdot \exp\left(-\frac{(x-x_c)^2}{2 w_2^2} \right) \ , & x \geq x_c 
\end{cases} 
$$
\noindent where $x_c$ is the peak centre, $H$ is the peak height, and $w_1$ and $w_2$ are the FWHM widths associated with the Gaussian curves left and right of the peak centre, respectively.

\subsection{Computational} \label{computational-methods}

Theoretical study of the dissociation mechanisms of the dication and trication of the fluorene was performed using an augmented Born-Oppenheimer molecular dynamics (BOMD) approach, described in detail in the ESI\dag. \cite{qceims1,qceims2,qceims_gfn,bbpp} Briefly, the excited electronic states are represented \textit{via} an "internal excess energy (IEE)/external energy reservoir" model, which operates as follows: the molecules interact with the laser pulses and are assigned an IEE using a recently proposed semi-heuristic model. \cite{bbpp} The IEE is then redistributed into vibrational degrees of freedom with a rate given by the internal conversion (IC) constant. The IC constant is computed using a classical electron-nuclei collision model. \cite{bbpp} The potential energy surfaces were computed using the GFN2-xTB method\cite{gfn2xtb} from the XTB semi-empirical package,\cite{xtbRev} which has been shown to work well for molecular fragmentation dynamics. \cite{qceims_gfn,bbpp} The Python fRAgmentation Molecular Dynamics (PyRAMD) source code developed is available for download with a detailed description of the computational methods used in the PyRAMD manual. \cite{pyramdCode,pyramdMan} The fragmentation dynamics were computed for fluorene with a time step of 1 fs and covering a total time of 10 ps in each BOMD trajectory. The XUV, or XUV and IR, pulses were applied simultaneously at the first step of the simulation. Simulations were performed over 100,000 CPU hours for the fluorene ions C$_{13}$H$_{10}^{2+}$, C$_{13}$H$_{10}^{3+}$, C$_{13}$T$_{10}^{3+}$ (i.e.~tritium substituted), with 13,000, 6,000 and 14,000 trajectories, respectively. C$_{13}$T$_{10}^{3+}$ was used to increase the number of trajectories leading to dissociation of the carbon backbone and is discussed in more detail in Section \ref{dication-section}. Vertical ionization potentials for the PAHs were calculated at the $\omega$B97/def2-TZVPP level of theory.

\section{Results and Discussion}

\subsection{Mass Spectrum}\label{mass-spectrum}

From the MCP readout, the TOF spectra for the PAH molecules were measured and converted to mass spectra, as shown in Fig. \ref{fig:mass-spectrum}. Due to the relatively high masses investigated and the time resolution of the spectrometer and ADC, hydrogen loss cannot be distinguished in the mass spectrum. Consequently, analysis of this data set revolves around the fragmentation of the carbon framework with the nomenclature C$_2$H$_x^+$ and C$_{11}$H$_y^+$ referring to the family of fragment ions containing two and eleven carbons atoms, respectively, with an undetermined number of hydrogen atoms. The time coordinate in the PImMS data set was converted to mass and matched against the mass spectrum from the MCP readout to allow ion images to be extracted for the various species of interest. 

Our mass spectra (Fig. \ref{fig:mass-spectrum}) show that upon interaction with an XUV photon, the primary products are the parent monocation and dication, with varying levels of hydrogen loss. These closely resemble previous experiments on PAHs in an ultrafast NIR regime with intensities of the order of 10$^{15}$ W cm$^{-2}$. \cite{ledingham1999multiply} In both experiments, the manifold of excited states is populated nearly instantaneously and electronic relaxation to form a stable parent ion apparently out-competes fragmentation of the carbon backbone. The non-adiabatic electronic relaxation of PAHs on the femtosecond timescale has been demonstrated in a number of recent laboratory experiments. \cite{lee2021time, marciniak2015xuv, herve2021ultrafast, boyer2021ultrafast} 

Previous experiments on aromatic molecules using nanosecond pulse length UV laser beams have demonstrated differences in the ratio of parent ions and fragment ions, depending on both the photon wavelength and the intensity.\cite{yang1983wavelength, dietz1982model, boesl1991multiphoton}  Laser intensities up to 10$^7$ W cm$^{-2}$ generally yield only parent ions in a "soft ionization" regime, whereas up to 10$^9$ W cm$^{-2}$, small fragments dominate the mass spectrum. The higher beam intensity allows the molecule to access more dissociative states following ionization through multiple photon absorption. The Keldysh parameter, which provides a metric for strong field ionization compared to ionization through multi-photon absorption, is in excess of 1 in these UV experiments indicating that the majority of ionization takes place \textit{via} multi-photon absorption, compared to a value of \textasciitilde 0.1 in the higher intensity NIR experiments. \cite{keldysh1965ionization}.

Using FLU as an example in the expanded view in Fig. \ref{fig:mass-spectrum}, it is clear that other than PAH$^+$ and PAH$^{2+}$, multiple fragment ion species are formed in lower abundances. Notable peaks include $m/z$ = 139 and 70, which are primarily attributed to the loss of neutral acetylene from the parent monocation and dication to form C$_{11}$H$_x^+$ and C$_{11}$H$_x^{2+}$, respectively (loss of the acetylene ion also contributes to these channels). Minor peaks corresponding to other dications, such as C$_7$H$_x^{2+}$, can be seen. Dications with an even number of carbons have a similar time-of-flight to smaller monocations, i.e., C$_{8}$H$_x^{2+}$ overlaps with C$_{4}$H$_y^{+}$, and therefore cannot be distinguished in the mass spectrum; however, contributions can be separated using covariance analysis, which is described in the following sections. Narrow peaks corresponding to stable PAH$^{3+}$ ions can be seen in all mass spectra, for example at $m/z$ $\approx$ 55 for the FLU$^{3+}$ ion. The ionization potentials for PAH$^{3+}$ for FLU, PHE, and PYR were calculated at the $\omega$B97/def2-TZVPP level of theory to be 38.4 eV, 37.8 eV, and 36.3 eV, respectively, which are all accessible with a single 40.9 eV XUV photon used in our experiments. Molecular nitrogen (N$_2^+$) from background gas was also seen to overlap with the C$_2$H$_x^+$ ion for all molecules and is discussed below.

\subsection{Dication Dissociation}\label{dication-section}

\begin{figure}[ht!]
\centering
\includegraphics[scale=0.62]{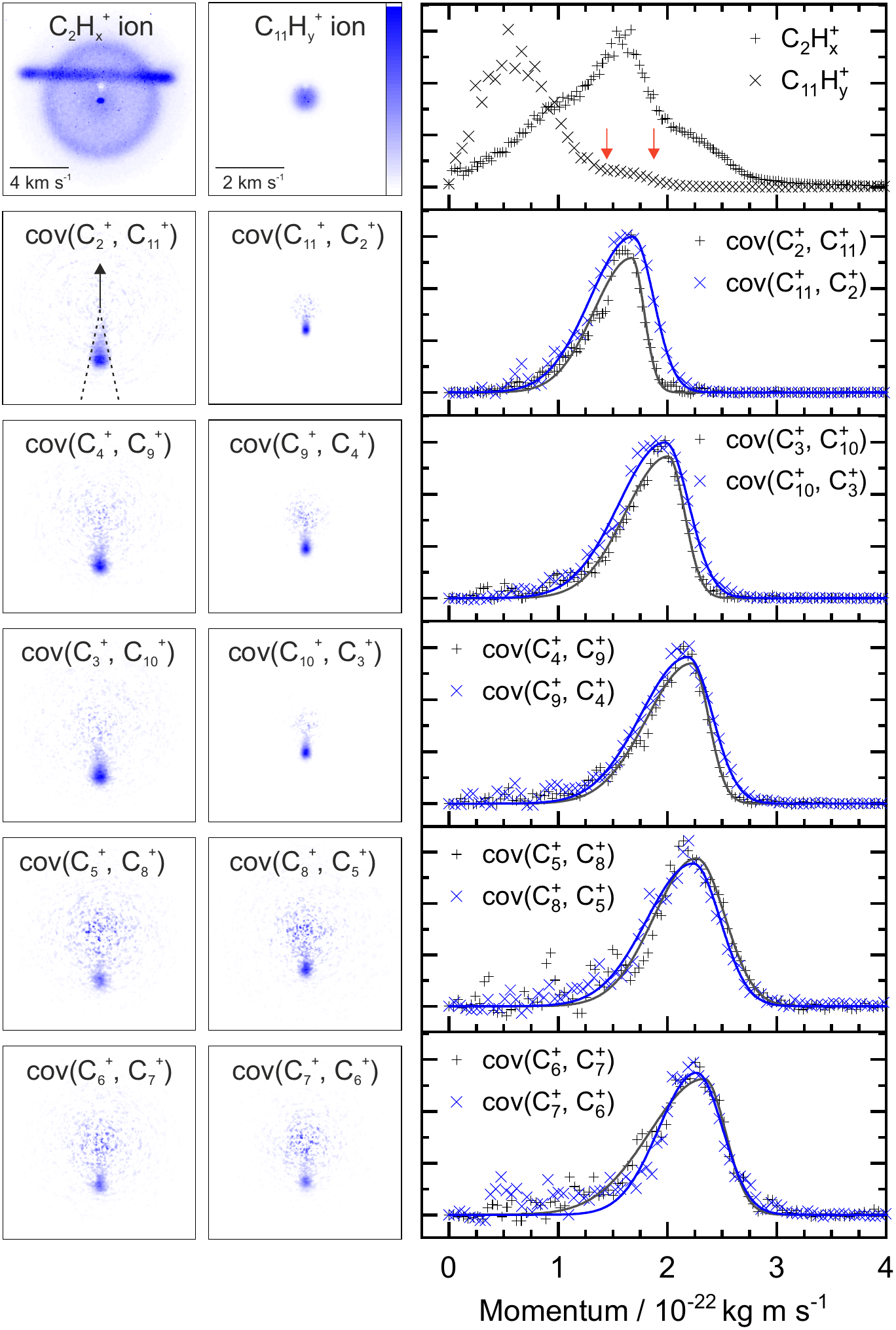}
\caption{\label{fig:flu2+dissociation} Top row, left: C$_2$H$_x^+$ and C$_{11}$H$_y^+$ ion images from FLU presented without image correction or symmetrization. Significant contributions from background N$_2^+$ can be seen in the C$_2$H$_x^+$ ion image. Between the red arrows indicate a minor shoulder in the C$_{11}$H$_y^+$ momentum distribution that is isolated in the covariance analysis. Top row, right: the momentum distributions extracted from the ion images. Remaining rows, left: the covariance images for ion pair dissociation from FLU$^{2+}$ with the nomenclature cov(A,B) referring to the covariance of ion A with the ion B as a reference, hydrogen atoms omitted for clarity. The arrow in cov(C$_2^+$, C$_{11}^+$) indicates the reference direction for all covariance map images. Remaining rows, right: the momentum profiles extracted from the covariance images using the region marked by dashed lines in the cov(C$_2^{+}$,C$_{11}^{2+}$) image. The recoil frame covariance analysis enables us to extract the pathway forming C$_2$H$_x^+$ and C$_{11}$H$_y^+$ from the C$_2$H$_x^+$ ion image, despite significant N$_2^+$ background ions. All images and momentum profiles individually normalized.}
\end{figure}

Fig. \ref{fig:flu2+dissociation} shows two example ion images acquired in the experiment, C$_2$H$_x^+$ and C$_{11}$H$_y^+$ from FLU. The C$_2$H$_x^+$ ion image contains contributions from all the dissociation channels creating the C$_2$H$_x^+$ ion, i.e., from the monocation, dication and trication, as well as significant contributions from the N$_2^+$ accounting for the horizontal line and the central spot. The normalized momentum distributions acquired from the ion images are shown on the right of Fig. \ref{fig:flu2+dissociation}. In the dissociative pathway C$_{13}$H$_{10}^{2+}$ $\rightarrow$ C$_{2}$H$_x^+$ + C$_{11}$H$_y^+$, the resulting ion should be momentum matched, and evidence for this can be vaguely seen at $\sim$1.6 $\times$ 10$^{-22}$ kg m s$^{-1}$ with the major peak in the C$_{2}$H$_x^{+}$ plot matching a small shoulder in the C$_{11}$H$_y^{+}$ equivalent (between the red arrows). The majority of C$_{11}$H$_y^+$ ions have a low momentum (below 1.25 $\times$ 10$^{-22}$ kg m s$^{-1}$) and are presumably produced by neutral C$_{2}$H$_x$ loss from the monocation (C$_{13}$H$_{10}^{+}$ $\rightarrow$ C$_{2}$H$_x$ + C$_{11}$H$_y^+$). This channel almost obscures the small shoulder in the C$_{11}$H$_y^+$ plot, and the C$_{2}$H$_x^+$ ion has a very broad contribution in the momentum distribution, owing to the line of N$_2^+$.

By plotting the recoil-frame covariance images for the ion pairs, shown in the second row in Fig. \ref{fig:flu2+dissociation}, the C$_2$H$_x^+$ signal produced in the same dissociative pathways as C$_{11}$H$_x^+$ can be isolated extremely effectively, despite the heavy N$_2^+$ contamination in the original C$_2$H$_x^+$ ion image. The momentum profiles from the region between the dashed lines in the cov(C$_2^+$, C$_{11}^+$) image are plotted to the right of the covariance images. Comparing the momentum profiles of the ion images and the covariance images (top and second row, respectively)  allows the contribution from the dicationic dissociation pathway to be distinguished unambiguously at the previously mentioned peaks at $\sim$1.6 $\times$ 10$^{-22}$ kg m s$^{-1}$.

The covariance map images and momentum profiles from the other two-body dissociation of the FLU$^{2+}$ ion are shown in the other rows of Fig. \ref{fig:flu2+dissociation}, each demonstrating an excellent match in momentum. The momentum values and corresponding kinetic energy values are extracted for the three PAHs and shown in \mbox{Table \ref{tab:PAH2+table}}, with the total kinetic energy release (TKER) calculated using:
\begin{equation} \label{eq:2}
\textrm{TKER} = \frac{m_1v_1^2}{2} + \frac{m_2v_2^2}{2}
\end{equation}
\noindent{}The agreement in momentum between ions resulting from the same dication is expected due to conservation of momentum:
\begin{equation} \label{eq:1}
|m_1v_1| = |m_2v_2|
\end{equation}
\noindent{}The covariance map images provide high precision in determining the ion momenta, and having two measurements of the momentum yields a reliable TKER value. Across the three PAHs studied, our experiments determined the dissociation energies of the parent dication to be in the range 1.94-2.60 eV.

\begin{table}
\caption{\label{tab:PAH2+table} Experimentally determined momenta of fragment ions from PAH dication dissociation from the covariance images and associated TKER. Errors listed for the momenta are the fit errors.}
\centering
\begin{adjustbox}{width=0.48\textwidth}
\tiny
\begin{tabular}{@{}cccccc@{}}
\toprule
                            & Ion 1                & Ion 2                & Momentum 1            & Momentum 2           & TKER            \\
\multicolumn{1}{l}{}        & \multicolumn{1}{l}{} & \multicolumn{1}{l}{} & \multicolumn{2}{c}{10$^{-22}$ kg m s$^{-1}$} & eV              \\ \midrule
\multirow{5}{*}{FLU$^{2+}$} & C$_{2}$H$_{x}^{+}$   & C$_{11}$H$_{y}^{+}$  & 1.67 $\pm$ 0.01       & 1.68 $\pm$ 0.02      & 2.40 $\pm$ 0.04 \\
                            & C$_{3}$H$_{x}^{+}$   & C$_{10}$H$_{y}^{+}$  & 2.01 $\pm$ 0.01       & 1.98 $\pm$ 0.02      & 2.51 $\pm$ 0.04 \\
                            & C$_{4}$H$_{x}^{+}$   & C$_{9}$H$_{y}^{+}$   & 2.22 $\pm$ 0.02       & 2.19 $\pm$ 0.02      & 2.60 $\pm$ 0.04 \\
                            & C$_{5}$H$_{x}^{+}$   & C$_{8}$H$_{y}^{+}$   & 2.26 $\pm$ 0.02       & 2.23 $\pm$ 0.02      & 2.47 $\pm$ 0.04 \\
                            & C$_{6}$H$_{x}^{+}$   & C$_{7}$H$_{y}^{+}$   & 2.34 $\pm$ 0.02       & 2.26 $\pm$ 0.02      & 2.46 $\pm$ 0.04 \\ \midrule
\multirow{7}{*}{PYR$^{2+}$} & C$_{2}$H$_{x}^{+}$   & C$_{14}$H$_{y}^{+}$  & 1.64 $\pm$ 0.02       & 1.62 $\pm$ 0.04      & 2.23 $\pm$ 0.07 \\
                            & C$_{3}$H$_{x}^{+}$   & C$_{13}$H$_{y}^{+}$  & 1.99 $\pm$ 0.03       & 2.05 $\pm$ 0.04      & 2.44 $\pm$ 0.07 \\
                            & C$_{4}$H$_{x}^{+}$   & C$_{12}$H$_{y}^{+}$  & 2.26 $\pm$ 0.03       & 2.17 $\pm$ 0.04      & 2.49 $\pm$ 0.07 \\
                            & C$_{5}$H$_{x}^{+}$   & C$_{11}$H$_{y}^{+}$  & 2.34 $\pm$ 0.05       & 2.33 $\pm$ 0.04      & 2.39 $\pm$ 0.09 \\
                            & C$_{6}$H$_{x}^{+}$   & C$_{10}$H$_{y}^{+}$  & 2.49 $\pm$ 0.04       & 2.33 $\pm$ 0.04      & 2.38 $\pm$ 0.08 \\
                            & C$_{7}$H$_{x}^{+}$   & C$_{9}$H$_{y}^{+}$   & 2.34 $\pm$ 0.04       & 2.37 $\pm$ 0.04      & 2.13 $\pm$ 0.08 \\
                            & C$_{8}$H$_{x}^{+}$   & C$_{8}$H$_{y}^{+}$   & 2.37 $\pm$ 0.01       & 2.37 $\pm$ 0.01      & 2.10 $\pm$ 0.03 \\ \midrule
\multirow{6}{*}{PHE$^{2+}$} & C$_{2}$H$_{x}^{+}$   & C$_{12}$H$_{y}^{+}$  & 1.64 $\pm$ 0.01       & 1.45 $\pm$ 0.02      & 2.21 $\pm$ 0.03 \\
                            & C$_{3}$H$_{x}^{+}$   & C$_{11}$H$_{y}^{+}$  & 1.94 $\pm$ 0.01       & 1.81 $\pm$ 0.02      & 2.27 $\pm$ 0.04 \\
                            & C$_{4}$H$_{x}^{+}$   & C$_{10}$H$_{y}^{+}$  & 2.13 $\pm$ 0.02       & 1.97 $\pm$ 0.02      & 2.27 $\pm$ 0.04 \\
                            & C$_{5}$H$_{x}^{+}$   & C$_{9}$H$_{y}^{+}$   & 2.20 $\pm$ 0.02       & 2.10 $\pm$ 0.02      & 2.18 $\pm$ 0.04 \\
                            & C$_{6}$H$_{x}^{+}$   & C$_{8}$H$_{y}^{+}$   & 2.15 $\pm$ 0.02       & 2.05 $\pm$ 0.02      & 1.94 $\pm$ 0.04 \\
                            & C$_{7}$H$_{x}^{+}$   & C$_{7}$H$_{y}^{+}$   & 2.17 $\pm$ 0.11       & 2.17 $\pm$ 0.11      & 2.02 $\pm$ 0.21 \\ \bottomrule
\end{tabular}
\end{adjustbox}
\end{table}

\begin{figure}
\centering
\includegraphics[scale=0.55]{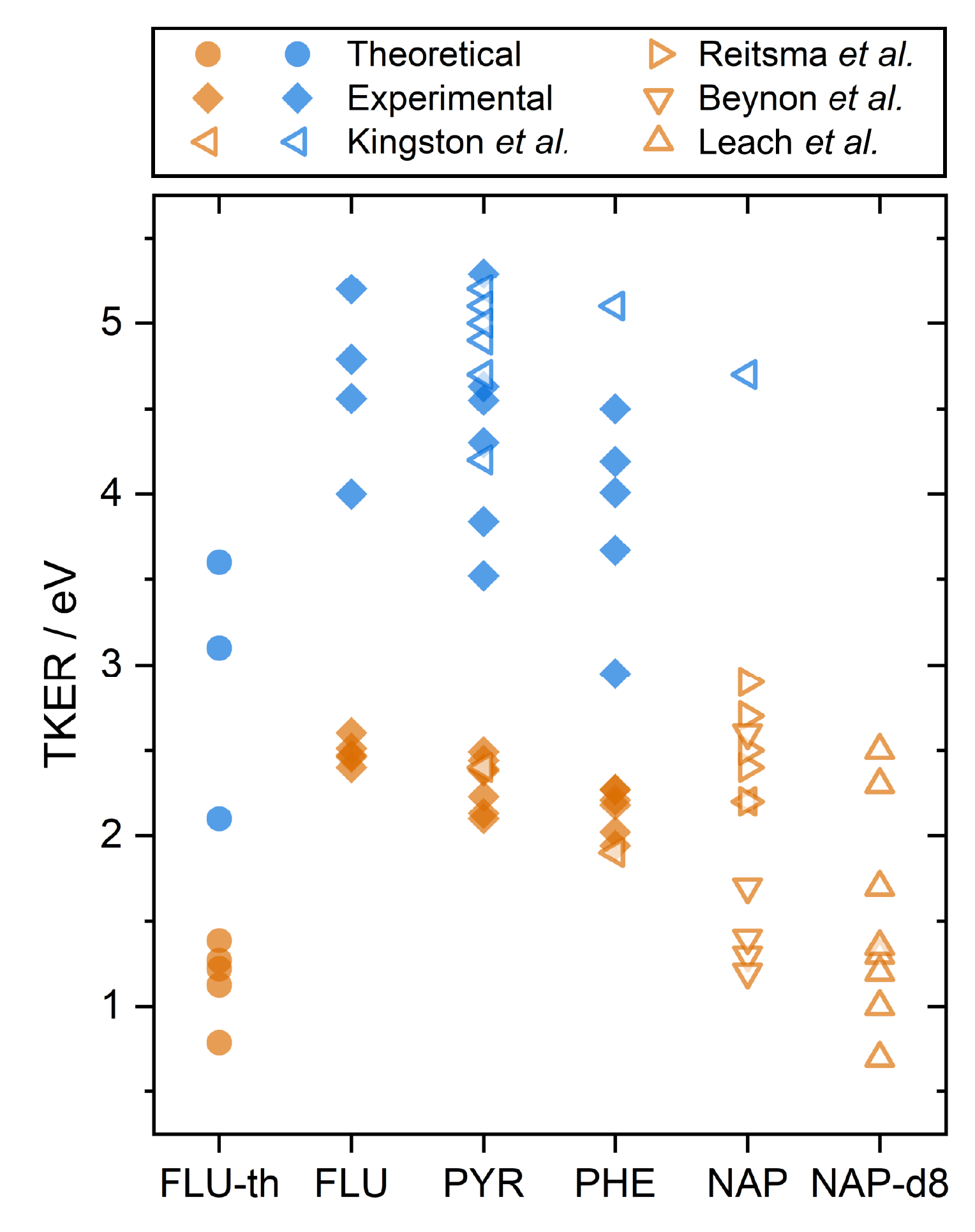}
\caption{\label{fig:KE-summary} A summary of PAH$^{2+}$ and PAH$^{3+}$ two-body dissociation TKERs for the various carbon backbone fragmentations. FLU-th refers to the theoretically determined values. Values from this study are shown in solid symbols, previous studies are shown with hollow symbols. \cite{kingston1985multiple, reitsma2013ion, beynon1970ion, leach1989formation} Orange and blue refer to dissociation from the PAH dication and trication, respectively.}
\end{figure}

Simulations of the fluorene dication (C$_{13}$H$_{10}^{2+}$) were performed as described in Section \ref{computational-methods}. The results are summarized alongside the experimental results in Fig. \ref{fig:KE-summary}. Briefly, the dication dissociation, C$_{13}$H$_{10}^{2+}$ $\rightarrow$ C$_n$H$_x^+$ + C$_{13-n}$H$_y^+$ (n = 2-6) is well characterized, predicting mean TKER values of 0.79-1.38 eV. These values qualitatively match the low experimental TKER values for FLU$^{2+}$ dissociation, although they are consistently \textasciitilde1.5 eV lower. This is attributed to two factors:

\begin{enumerate}
    \item The BOMD method employed predicts dissociation only from the electronic ground state. In reality, electronically excited states will contribute to dissociation and will lead to a higher TKER.
    \item A semi-empirical approach was used to compute the highly energetic conformations of the molecules, which could lead to systematic errors in the potential.
\end{enumerate}

\noindent With the above considerations, the computed values are expected to underestimate the TKER values. 

A comparison can also be made to previous dication TKER measurements on PYR and PHE, as well as naphthalene (NAP, C$_{10}$H$_{8}$) and NAP-d8 (C$_{10}$D$_{8}$), shown by the orange symbols in Fig. \ref{fig:KE-summary}. \cite{kingston1985multiple, reitsma2013ion, beynon1970ion, leach1989formation} As described in the introduction, the experiments by Kingston \textit{et al.}~utilised 70 eV electron impact and IKES. For PYR and PHE, they reported the dissociation energies for one dissociation channel each (C$_3$H$_3^+$ and C$_2$H$_2^+$ loss, respectively), which are in remarkable good agreement with our experimental values. NAP and NAP-d8 provide other interesting comparison points, particularly because NAP-d8$^{2+}$ measurements by Leach \textit{et al}.~used rare gas lamps at the same wavelength as the present study. Looking first at the NAP values, a wide range (1-3 eV) is reported across the three data sets. It is notable that the measurements by Beynon \textit{et al.}~are skewed towards the lower end of the energy range, which might indicate a systematic underestimation. Comparing our FLU, PYR, and PHE dication TKER values to the NAP$^{2+}$ dissociation measurements by Reitsma \textit{et al}., the TKER values across the molecules are relatively similar, with the Reitsma values being slightly higher on average. In their experiments, NAP$^{2+}$ is formed through impact with a 30 keV He$^{2+}$ ion instead of photoionization, which would likely populate a different ensemble of states. Higher TKER values could result from more internal energy in the dication, and consequently a more impulsive dissociation. The density of states is also lower for smaller molecules, resulting in higher microcanonical vibrational excitation, which may increase the TKER. Further, Coulombic repulsion between fragments is likely to be stronger for NAP than the PAHs in the present study due to being a smaller molecule. NAP-d8$^{2+}$ measurements by Leach \textit{et al}.~recorded TKER values for two-body dissociation in the range 0.7-2.5 eV. This is a relatively large range for a single study and reports the lowest experimental TKER values of any study. Fully deuterating the molecule increases the molecular mass (136 vs 128 a.m.u.) and reduces the vibrational energy level spacing, which could significantly increase the rate of internal relaxation, and therefore promote dissociation from lower electronic states. This might also explain why there is such a good match with the theoretical results for FLU, which have been calculated assuming the electronic ground state. Comparing the TKER values of NAP and NAP-d8 using the recoil-frame covariance map imaging techniques in this paper would form an interesting basis for a future investigation.

Our experimental TKER results for PAH$^{2+}$ dissociation are in the range 1.94-2.60 eV. With simple molecules, such as diatomics or substituted methane molecules, or at very high charge states where the Coulombic effects are dominant, the ionic fragments are often approximated as hard spheres with instantaneous dissociation. In doing so, the distance between the fragments, \textit{r}, can be calculated by assuming that the TKER arises from conversion of the electric potential energy ($U_E$) into kinetic energy:

\begin{equation}
    TKER = U_E(r) = k_e\frac{q_1q_2}{r}
\end{equation}

\noindent where $k_e$ refers to Coulomb's constant. Using such a model, some of our TKER values would be consistent with a minimum distance greater than 7 \r{A}, which is approximately the distance between the furthest carbon atoms in a pyrene molecule, so such a simple model is clearly not sufficient for the PAH molecules. Fragmentation of the carbon backbone requires breaking at least two $\sigma$ bonds and overcoming the stability of the aromatic system, likely accompanied by significant molecular rearrangement. This is consistent with our BOMD simulations, which predict PAH isomerization including ring opening and elongation into chains. Several of the comparative experimental studies discussed in this section also postulated significant molecular rearrangements prior to dissociation, and a number of theoretical studies have anticipated similar pathways accessible within a few eV. \cite{solano2015complete, trinquier2017pah, trinquier2017pah2, west2014photodissociation} The range of TKER values measured may be thought of as being low considering that the double ionization potential of the PAHs studied is approximately 20 eV, \cite{holm2011dissociation} so formation of a dication through absorption of a 40.9 eV photon should result in 0-20 eV of internal energy (with remaining energy imparted into the ejected electrons). This is because the channels in this study are two-body fragmentations which result from the lower lying PAH$^{2+}$ states. PAH$^{2+}$ ions that are formed with higher internal energy would be expected to undergo more extensive fragmentation into three or more bodies, which was also reproduced in our simulations.

\subsection{Trication Dissociation}\label{trication-section}

\begin{figure}
\includegraphics[scale=0.62]{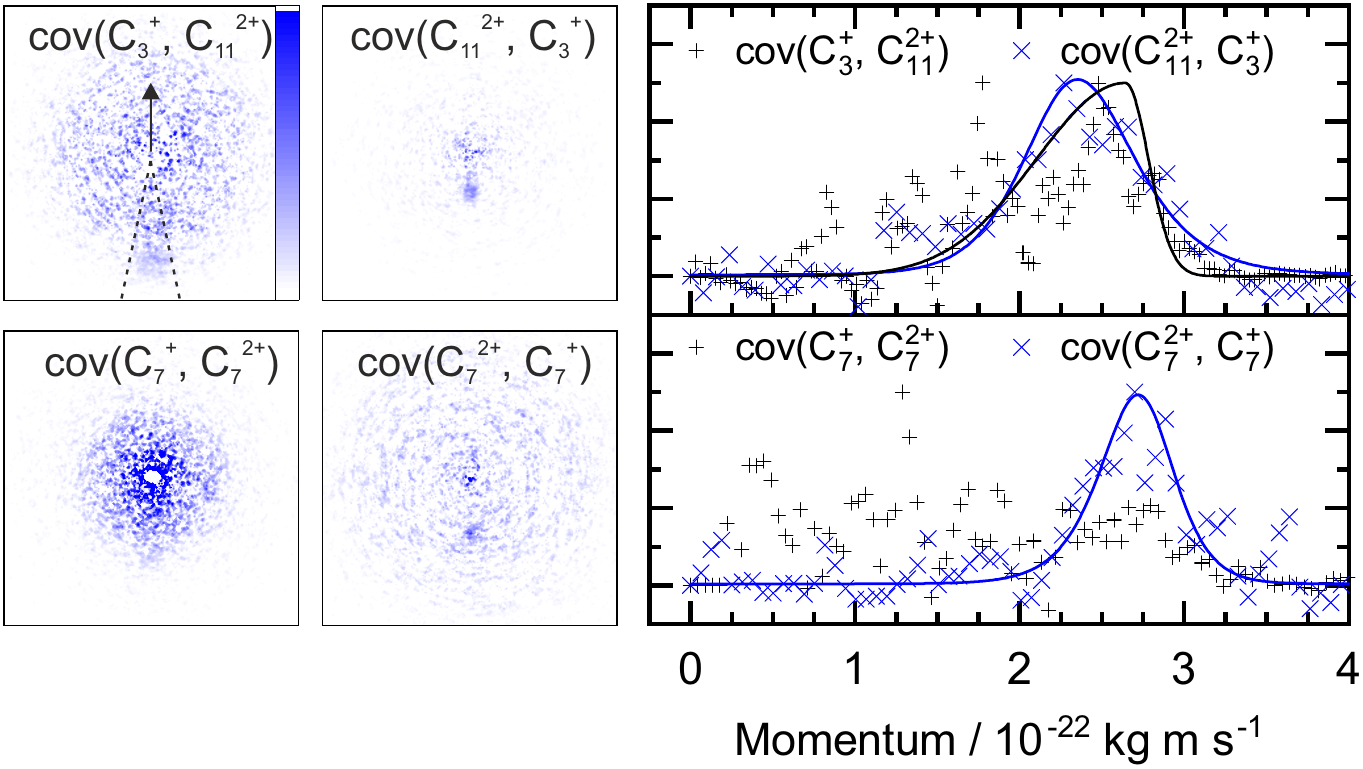}
\caption{\label{fig:flu3+dissociation} Left: the covariance map images for dissociation of PHE$^{3+}$ into C$_3$H$_x^+$ and C$_{11}$H$_y^{2+}$, and C$_7$H$_x^+$ and C$_{7}$H$_y^{2+}$, top and bottom, respectively. Nomenclature as in Fig. \ref{fig:flu2+dissociation}. Covariance signal in the cov(C$_7^+$,C$_{7}^{2+}$) image cannot be discerned above the noise level but the cov(C$_7^{2+}$,C$_{7}^{+}$) image provides the TKER for the dissociation process. The centre of the cov(C$_7$H$_x^+$,C$_{7}$H$_y^{2+}$) image shows detector saturation due to the flight times of C$_7$H$_x^+$ and C$_{14}$H$_y^{2+}$ overlapping. Right: the momentum profiles extracted from the corresponding covariance images using the region marked by dashed lines in the cov(C$_3^{+}$,C$_{11}^{2+}$) image. All images and momentum profiles are individually normalized.}
\end{figure}

A similar analysis can be performed on the ions resulting from dissociation of the parent trication. This poses a number of challenges. Firstly, the signal level is significantly lower than from dication dissociation owing to the fact that the 40.9 eV photons are only a few eV above energy required to triply ionize these molecules. \cite{lee2021time, kingston1985multiple} Secondly, two-body dissociation of the parent trication will result in a monocation and a dication. As mentioned in Section \ref{mass-spectrum}, dications containing an even number of carbons overlap with monocations in flight time due to having an equivalent value of $m/z$, for example, C$_6$H$_x^{2+}$ has the same arrival time, and is therefore recorded in the same ion image as C$_3$H$_y^{+}$. Despite these constraints, clear signal can be seen in the covariance map images for a number of ion pairs. Example covariance maps from PHE$^{3+}$ dissociation are shown in Fig. \ref{fig:flu3+dissociation} where the signal can be seen in the cov(C$_{11}$H$_x^{2+}$, C$_3$H$_y^+$) and cov(C$_3$H$_x^+$, C$_{11}$H$_y^{2+}$) images. Where covariance map images do not show a clear signal, complete information about the dissociation process can still be extracted provided that one of the covariance map images in an ion pair gives the momentum. This can be seen by combining equations (\ref{eq:2}) and (\ref{eq:1}) to yield:

\begin{equation}
    TKER = \frac{m_1v_1^2}{2} + \frac{m_2v_2^2}{2}  = \frac{m_1v_1^2}{2} + \frac{m_1^2v_1^2}{2m_2}
\end{equation}

\noindent{}For instance, the second row of Fig. \ref{fig:flu3+dissociation} shows the covariance map plots associated with the pathway C$_{14}$H$_{10}^{3+}$ $\rightarrow$ C$_{7}$H$_x^+$ + C$_{7}$H$_y^{2+}$. In the cov(C$_7$H$_x^+$,C$_{7}$H$_x^{2+}$) momentum profile, no suitable peak can be discerned above the noise level due to a combination of low signal levels for this pathway and the fact that the C$_7$H$_x^+$ ion image overlaps in time-of-flight with C$_{14}$H$_y^{2+}$, causing detector saturation and non-quantitative measurement of the ion image. In contrast, the corresponding cov(C$_{7}$H$_x^{2+}$, C$_7$H$_x^+$) image has a clear covariance point and peak in the covariance map and the momentum plot, respectively. Using the same momentum value for C$_{7}$H$_x^{2+}$ and C$_{7}$H$_x^{+}$ (due to conservation of momentum in the dissociation process) allows the corresponding TKER to be calculated. Momentum and TKER values are reported in Table \ref{tab:PAH3+table} covering the range 2.95-5.29 eV for the PAH trication dissociation.

\begin{table}
\caption{\label{tab:PAH3+table} Experimentally determined momenta of fragment ions from PAH trication dissociation from the covariance images and associated TKER. Errors listed for the momenta are the fit errors.}
\centering
\begin{adjustbox}{width=0.48\textwidth}
\tiny
\begin{tabular}{@{}cccccc@{}}
\toprule
                            & Ion 1                & Ion 2                & Momentum 1            & Momentum 2           & TKER            \\
\multicolumn{1}{l}{}        & \multicolumn{1}{l}{} & \multicolumn{1}{l}{} & \multicolumn{2}{c}{10$^{-22}$ kg m s$^{-1}$} & eV              \\ \midrule
\multirow{5}{*}{FLU$^{3+}$} & C$_{2}$H$_{x}^{+}$   & C$_{11}$H$_{y}^{2+}$ & 2.47 $\pm$ 0.03       & 2.41 $\pm$ 0.05      & 5.20 $\pm$ 0.11 \\
                            & C$_{3}$H$_{x}^{+}$   & C$_{10}$H$_{y}^{2+}$ & -                     & 2.76 $\pm$ 0.09      & 4.79 $\pm$ 0.32 \\
                            & C$_{4}$H$_{x}^{+}$   & C$_{9}$H$_{y}^{2+}$  & 2.96 $\pm$ 0.04       & 2.86 $\pm$ 0.06      & 4.56 $\pm$ 0.13 \\
                            & C$_{5}$H$_{x}^{+}$   & C$_{8}$H$_{y}^{2+}$  & -                     & -                    & -               \\
                            & C$_{6}$H$_{x}^{+}$   & C$_{7}$H$_{y}^{2+}$  & 2.89 $\pm$ 0.08       & 2.99 $\pm$ 0.06      & 4.00 $\pm$ 0.20 \\ \midrule
\multirow{6}{*}{PYR$^{3+}$} & C$_{2}$H$_{x}^{+}$   & C$_{14}$H$_{y}^{2+}$ & 2.35 $\pm$ 0.05       & 2.43 $\pm$ 0.07      & 4.63 $\pm$ 0.19 \\
                            & C$_{3}$H$_{x}^{+}$   & C$_{13}$H$_{y}^{2+}$ & 2.96 $\pm$ 0.05       & 2.88 $\pm$ 0.08      & 5.29 $\pm$ 0.20 \\
                            & C$_{4}$H$_{x}^{+}$   & C$_{12}$H$_{y}^{2+}$ & -                     & 2.94 $\pm$ 0.10      & 4.30 $\pm$ 0.31 \\
                            & C$_{5}$H$_{x}^{+}$   & C$_{11}$H$_{y}^{2+}$ & 3.19 $\pm$ 0.20       & 3.32 $\pm$ 0.68      & 4.55 $\pm$ 1.09 \\
                            & C$_{6}$H$_{x}^{+}$   & C$_{10}$H$_{y}^{2+}$ & -                     & 2.96 $\pm$ 0.14      & 3.52 $\pm$ 0.35 \\
                            & C$_{7}$H$_{x}^{+}$   & C$_{9}$H$_{y}^{2+}$  & 3.08 $\pm$ 0.10       & 3.27 $\pm$ 0.06      & 3.84 $\pm$ 0.20 \\ \midrule
\multirow{6}{*}{PHE$^{3+}$} & C$_{2}$H$_{x}^{+}$   & C$_{12}$H$_{y}^{2+}$ & -                     & 2.30 $\pm$ 0.09      & 4.50 $\pm$ 0.37 \\
                            & C$_{3}$H$_{x}^{+}$   & C$_{11}$H$_{y}^{2+}$ & 2.64 $\pm$ 0.07       & 2.46 $\pm$ 0.09      & 4.19 $\pm$ 0.25 \\
                            & C$_{4}$H$_{x}^{+}$   & C$_{10}$H$_{y}^{2+}$ & -                     & 2.65 $\pm$ 0.17      & 3.67 $\pm$ 0.50 \\
                            & C$_{5}$H$_{x}^{+}$   & C$_{9}$H$_{y}^{2+}$  & 2.94 $\pm$ 0.07       & -                    & 4.01 $\pm$ 0.19 \\
                            & C$_{6}$H$_{x}^{+}$   & C$_{8}$H$_{y}^{2+}$  & -                     & -                    & -               \\
                            & C$_{7}$H$_{x}^{+}$   & C$_{7}$H$_{y}^{2+}$  & -                     & 2.63 $\pm$ 0.06      & 2.95 $\pm$ 0.13 \\ \bottomrule
\end{tabular}
\end{adjustbox}
\end{table}

Simulations of the fluorene trication (C$_{13}$H$_{10}^{3+}$) were performed as described in Section \ref{computational-methods}. From the fluorene trication (C$_{13}$H$_{10}^{3+}$), the vast majority of trajectories resulted in the immediate loss of a H$^+$ or H$_2^+$ ion. The resulting C$_{13}$H$_{8/9}^{2+}$ ions behave almost identically to the fluorene dication (C$_{13}$H$_{10}^{2+}$), further fragmenting along the carbon backbone to produce two monocations. Given that H$^+$ or H$_2^+$ loss has little KER associated with it, the associated trication dissociation TKER is very similar to that of the dication. To reduce the trajectories leading to loss of H$^+$ or H$_2^+$, and therefore increase the number leading to carbon backbone fragmentation, the hydrogen atoms were changed from the protium isotope to tritium in the simulations, creating the tritiated fluorene trication (C$_{13}$T$_{10}^{3+}$). The change in mass from 1 to 3 atomic units should have a negligible effect on the calculated TKER from fragmentation of the carbon framework. The C$_{13}$H$_{10}^{3+}$ and C$_{13}$T$_{10}^{3+}$ trajectories corresponding to dissociation pathways to those listed in Table \ref{tab:PAH3+table} predicted TKERs in the range 2.1-3.6 eV. These are consistently \textasciitilde2.0 eV below the experimental values for FLU$^{3+}$ dissociation (compared in Fig. \ref{fig:KE-summary}), which is attributed to the same reasons as in Section \ref{dication-section}, which would lead to systematic underestimation of the TKER. Similar to the results for the dication, the experimental and theoretical TKER values are generally found to be very low, attributed to molecular rearrangement and the requirement to overcome significant binding energy. Animations of dissociation of PAH$^{3+}$ from these simulations can be found in the ESI\dag.

Although trication dissociation has not been as widely studied as dication dissociation, Kingston \textit{et al.}~measured the TKER for a number of pathways for PYR$^{3+}$, which fall nicely within the range of our measurements, as shown by comparing the blue symbols in Fig. \ref{fig:KE-summary}. For PHE$^{3+}$, they reported only one value of 5.1 eV, which is 0.6 eV greater than the highest TKER value in the present study. This is curious given the match in the other measurements from Kingston \textit{et al.}, but not so different from our measurements to cause concern.

\section{Conclusions}
Using a combination of experimental and theoretical approaches, the single-photon XUV dissociative ionization of a series of small PAHs was investigated. In all cases, the mass spectra were found to consist primarily of the parent ion and dication, alongside lower quantities of fragment ions resulting from cleavage of the carbon backbone. Application of recoil-frame covariance isolated particular dissociative pathways, allowing the associated TKERs to be extracted. Ion pairs from PAH$^{2+}$ were found to have an excellent match in momentum, confirming the validity of this analysis method. Applying the same method to the PAH trication produced reliable TKER values, including dissociation pathways where signal was clear in only one of the fragments. Previous applications of recoil-frame covariance have typically studied small molecules with a limited number of dissociative pathways, and/or molecules tagged with halogen atoms to identify particular fragmentation channels \cite{burt2018communication, allum2018coulomb, slater2014covariance, slater2015coulomb, lee2020three}. The results presented here show that information can be extracted from many-atom molecules, providing a valuable tool to investigate the chemical dynamics of larger systems.

The experimental and theoretical results for carbon backbone fragmentation were found to be below 2.60 eV and 5.29 eV for the dication and trication, respectively, indicating significant molecular rearrangement and residual binding in the dissociation process. Our results have demonstrated that, with a sufficient source of photon energy, the PAH molecules studied are able to form a wide variety of ions. This is interesting from an astrochemical perspective as the ions formed have the potential to act as building blocks for larger molecules in a "bottom-up" model. \cite{tielens2013molecular}  In addition, the dissociation energy of the fragment ions in our experiments would be sufficient to overcome substantial association barriers in PAH formation channels in the otherwise cold environment of the ISM. Given that larger PAHs (i.e.~more than 50 carbon atoms) have been found to be more stable with respect to dissociation, \cite{west2019large} a scenario exists where the fragmentation of small PAHs facilitates the growth of larger PAHs. This would be consistent with the "grandPAH hypothesis" which was postulated by Tielens in 2013 and speculates that the family of PAH molecules in photodissociation regions is dominated by a limited number of large, highly symmetric, and stable PAHs. \cite{tielens2013molecular} Since then, there have been a number of telescope observations of photodissociative regions and laboratory measurements corroborating this theory, continuing to expand our understanding of the chemistry of the ISM. \cite{andrews2015pah, shannon2015probing, bouwman2021mid}. 


\section*{Conflicts of interest}
There are no conflicts to declare.

\section*{Acknowledgements}
This work was supported by the ERC Starting Grant ASTROROT, grant number 638027, and the project CALIPSOplus under the grant agreement 730872 from the EU Framework Programme for Research and Innovation HORIZON 2020. The experimental parts of this research were carried out with beamtime allocated for proposal F-20170540 at beamline BL1 FLASH at DESY (Hamburg, Germany), a member of the Helmholtz Association HGF. We acknowledge the Max Planck Society for funding the development and the initial operation of the CAMP end-station within the Max Planck Advanced Study Group at CFEL and for providing this equipment for CAMP@FLASH. The installation of CAMP@FLASH was partially funded by the BMBF grants 05K10KT2, 05K13KT2, 05K16KT3 and 05K10KTB from FSP-302. We acknowledge financial support by the European Union’s Horizon 2020 research and innovation program under the Marie Skłodowska-Curie Grant Agreement 641789 “Molecular Electron Dynamics investigated by Intense Fields and Attosecond Pulses” (MEDEA), the Clusters of Excellence “Center for Ultrafast Imaging” (CUI, EXC 1074, ID 194651731), the “Advanced Imaging of Matter” (AIM, EXC 2056, ID 390715994) of the Deutsche Forschungsgemeinschaft (DFG), and the Helmholtz Gemeinschaft through the “Impuls-und Vernetzungsfonds”. Additionally, the project was supported by The Netherlands Organization for Scientific Research (NWO) and is part of the Dutch Astrochemistry Network (DAN) II (Project No. 648.000.029). S.M., J.L., J.P., and P.E.-J. acknowledge support from the Swedish Research Council and the Swedish Foundation for Strategic Research. The authors are additionally thankful for support from the following funding bodies: the UK EPSRC (M.Br.\ and C.V.\ - EP/L005913/1, EP/T021675/1, EP/V026690/1; M.Bu.\ - EP/S028617/1), STFC (PNPAS award and mini-IPS Grant No. ST/J002895/1); the National Science Foundation (D.Rol.\ - PHYS-1753324); and the Helmholtz Initiative and Networking Fund through the Young Investigators Group Program (S.B.). L.H. acknowledges the support by the National Natural Science Foundation of China (11704147) and a fellowship within the framework of the Helmholtz-OCPC postdoctoral exchange program. For the purpose of open access, the author has applied a CC BY public copyright license to any Author Accepted Manuscript version arising from this submission.



\balance


\bibliography{rsc} 
\bibliographystyle{rsc} 

\end{document}